\documentclass[runningheads]{llncs}

\usepackage[utf8]{inputenc}
\usepackage{bold-extra}
\usepackage{amsmath,
	    amsfonts,
	    amstext,
	    amssymb,
	    mathrsfs,
	    mathpartir,
	    stmaryrd,
	    latexsym,
	    enumitem,
	    proof,
	    graphicx,
	    color,
	    hyperref,
	    comment,
	    mathtools}

\usepackage{semantic}
\usepackage{float}
\usepackage{tikz}
\usetikzlibrary{arrows,positioning,arrows.meta}
\tikzset{modal/.style = {>= stealth', shorten >= 0pt, shorten <= 0pt, auto,
			 node distance = 1cm, semithick}, 
	 point/.style = {circle, draw, fill = black, inner sep = 0.5mm}}

\newcommand{\ms}[1]
	{\null\ifmmode\mathord{\mathcode`-="702D\it #1\mathcode`\-="2200}
	\else$\mathord{\mathcode`-="702D\it #1\mathcode`\-="2200}$\fi}

\newcommand{\cws}[2]
	{\\ \centerline{$#2$} \\[-#1pt]}

\newcommand{\bibtrick}[1]
	{}

\newcommand{\cala}
        {\mathcal{A}}

\newcommand{\realns}
	{\mathbb{R}}

\newcommand{\procs}
	{\mathbb{P}}

\newcommand{\auxarrow}
	{\mathop{\longrightarrow}}

\newcommand{\wauxarrow}
	{\mathop{= \!\!\!\! \Longrightarrow}}

\newcommand{\nil}
	{\underline 0}

\newcommand{\eqdef}
	{\triangleq}

\newcommand{\wbis}[1]
        {\approx_{#1}}

\newcommand{\pco}[1]
	{\mathop{\Vert_{#1}}}

\begin{document}

\title{Formal Modeling and Verification \\ of the Algorand Consensus Protocol in CADP}
\titlerunning{Formal Modeling and Verification of Algorand Consensus Protocol in CADP}

\author{Andrea Esposito\inst{1} \and Francesco P. Rossi\inst{1} \and \\
Marco Bernardo\inst{1} \and Francesco Fabris\inst{2} \and Hubert Garavel\inst{3}}
\authorrunning{A. Esposito \and F.P. Rossi \and M. Bernardo \and F. Fabris \and H. Garavel}

\institute{Dip.\ Scienze Pure e Applicate, Univ.\ Urbino, 61029 Urbino, Italy \and 
Dip.\ Matematica, Informatica e Geoscienze, Univ.\ Trieste, 34127 Trieste, Italy \and
Univ.\ Grenoble Alpes, \textsc{Inria}, \textsc{Cnrs}, Grenoble \textsc{Inp}, \textsc{Lig}, 38000 Grenoble,
France}

\maketitle


\begin{abstract}
Algorand is a scalable and secure permissionless blockchain that achieves proof-of-stake consensus via
cryptographic self-sortition and binary Byzantine agreement. In this paper we present a process algebraic
model of the Algorand consensus protocol with the aim of enabling formal verification. Our model captures
the behavior of participants in terms of the structured alternation of consensus steps toward a
committee-based agreement. We validate the correctness of the protocol in the absence of adversaries and
then extend our model to assess the influence of coordinated malicious nodes that can force the commit of an
empty block instead of the proposed one. The adversarial scenario is analyzed through an
equivalence-checking-based noninterference framework that we have implemented in the CADP verification
toolkit. In addition to highlighting both the robustness and the limitations of the Algorand protocol under
adversarial assumptions, this work illustrates the added value of using formal methods for the analysis of
consensus algorithms within blockchains.
\end{abstract}

%
%
\section{Introduction}
\label{sec:intro}
%
%

A \emph{blockchain} is a distributed, tamper-proof ledger system that permanently records transactions
across a network of possibly untrusted nodes. The ledger maintains a cryptographically linked list of
committed blocks. In addition to transactions, each block contains a reference to the previous block,
forming an immutable chain of records. In its pure form, called \emph{public} or \emph{permissionless}
blockchain, any node can join or leave the system at any time, with agreement on transaction history being
reached via a consensus protocol instead of a centralized trusted party. The growing adoption of blockchain
systems in many domains, such as decentralized finance, digital identity, and secure voting~\cite{CXSZZ18},
is making consensus protocols a central subject of theoretical and applied research.

The two primary paradigms for consensus in public blockchains are \emph{proof of work} (PoW) and \emph{proof
of stake} (PoS). The former, initially introduced for \emph{Bitcoin}~\cite{Nak08} and still used by several
distributed ledgers, requires participants to solve computationally intensive puzzles, which guarantees
security at the cost of scalability and high energy consumption. In contrast, PoS protocols, subsequently
adopted by blockchains such as \emph{Ethereum}~\cite{But14}, assign block validation rights proportionally
to staked tokens, thus significantly reducing energy consumption, improving throughput, and lowering entry
barriers for validators.

A notable proposal in this context is \emph{Algorand}~\cite{CM19}, a permissionless blockchain designed to
achieve scalability, decentralization, and security. Algorand introduces several innovations that set it
apart from other PoS blockchains. Its distinctive PoS consensus mechanism, based on \emph{cryptographic
self-sortition} and \emph{Byzantine agreement}~\cite{PSL80,Fis83,DS83,FLP85,Lyn96,Mul93}, enables Algorand
to reach consensus on blocks in a matter of seconds, even in the presence of powerful adversaries. Its
two-phase protocol structure, comprising \emph{graded consensus} (GC) followed by \emph{binary Byzantine
agreement} (BBA$^{*}$), with the latter being an adaptation to a public-key setting of the probabilistic
binary protocol of~\cite{FM97}, guarantees agreement and safety under minimal assumptions. It employs a
decentralized, stake-weighted committee selection process where nodes are chosen through \emph{verifiable
random functions} (VRFs)~\cite{MRV99} not only probabilistically but also privately -- so as to ensure that
committee members are not known at the beginning of a consensus step -- and change at each consensus step --
so as to avoid corruption.

Algorand stands out not only for the aforementioned technical innovations, but also as a significant player
in the blockchain ecosystem, with a current market capitalization of approximately 2.24 billion USD that
places it among the top 50 cryptocurrencies~\cite{CMC25}. Since its inception, Algorand has raised about 66
million USD in funding across three institutional rounds, backed by investors such as Union Square Ventures,
Pillar VC, NGC Ventures, and Lemniscap~\cite{CC25}. Moreover, Algorand is also active in the decentralized
finance sector, with a total value locked of 82.3 million USD today and a peak of nearly 320 million USD
reached in February 2024~\cite{DL25}. Together, these figures reflect both the confidence of the investment
community and the tangible scale of the network economic presence.

The complexity of Algorand consensus logic, in particular under adversarial conditions, makes formal
verification both challenging and necessary. The correctness of the protocol hinges on subtle properties
such as the behavior of randomized bit setting, committee overlap, and cryptographic credential validity,
all of which are difficult to reason about informally or through testing alone. Formal modeling allows for
unambiguous specification of protocol dynamics and enables the use of automated verification tools to prove
properties related to safety, liveness, security, and performance.

In this paper, we present a process algebraic model of the Algorand consensus protocol, with a specific
focus on the BBA$^{*}$ phase. We formalize the behavior of honest nodes and malicious nodes by using a
probabilistic process calculus and encode the protocol key dynamics related to bit propagation, step
transitioning, and committee-based decision making. Our model reflects the probabilistic and stake-weighted
nature of committee selection and abstracts the network under a \emph{fully synchronous} execution
assumption both for theoretical reasons and to facilitate analysis. Honest nodes are expected to follow the
protocol as specified, ensuring its correct execution. Malicious nodes, on the other hand, may attempt to
disrupt consensus by obstructing or boycotting a proposed block. We translate the model into \emph{LNT} for
verification with the \emph{CADP} toolkit~\cite{GLMS13}.

We analyze the protocol under benign and adversarial scenarios. In the presence of sufficiently many
malicious nodes that coordinate to boycott a proposed block, we show that the protocol is forced to commit
an empty block, as allowed by its design. To reason about this behavior, we adopt a noninterference-based
security framework~\cite{GM82,FG01}, define the adversarial coordination action as a high-level action, and
demonstrate, through equivalence checking, that its presence can observably affect the protocol output. This
not only confirms the limits of Algorand resilience, but also highlights the importance of formal methods
for understanding the guarantees and vulnerabilities of consensus protocols.

%
\subsection{Related Works}
\label{sec:related_works}
%

Formal modeling and verification of consensus protocols are receiving an increasing attention in the
blockchain community~\cite{VYC22}. As an example, communicating-automata-like models and the PRISM model
checker~\cite{KNP11} have been considered in~\cite{Ves23} to formally analyze consensus in Bitcoin and a
variant of Ethereum.

Some formal models of Algorand have been developed in different frameworks. A Coq formalization of the
Algorand consensus protocol has been provided in~\cite{ACLMOPR19} as a global state transition system,
accounting for adversarial behavior and timing assumptions and proving safety in an asynchronous setting. A
formal model of Algorand stateless smart contracts has been constructed in~\cite{BBLSZ21} by focusing on
properties such as value preservation and determinism. Although this work targets contract logic rather than
consensus, it underscores the applicability of formal methods across different components of the Algorand
ecosystem.

Like in our approach, other researchers have applied process algebra to model well-known consensus
protocols. A notable example is the formalization of the Raft protocol in LNT presented in~\cite{Evr20},
where the protocol is verified under scenarios involving node failures and leader re-elections by using the
CADP toolset. Our work offers the first process algebraic model of the BBA$^{*}$ phase of Algorand and
employs equivalence-checking-based noninterference as a formal technique to detect illicit behaviors by
malicious actors. The model supports automated verification of both honest-only executions and adversarial
scenarios.

%
\subsection{Paper Organization}
\label{sec:paper_org}
%

The remainder of this paper is structured as follows. Section~\ref{sec:background} provides a detailed
description of the Algorand consensus protocol, including the GC and BBA$^{*}$ phases, as well as the
underlying network assumptions. Section~\ref{sec:model_algorand} introduces our probabilistic process
algebra model for BBA$^{*}$, first under the assumption of fully honest participants and then extended to
include adversarial nodes. Section~\ref{sec:verification} explains how the model is translated into LNT,
details the chosen parameters, and presents the noninterference-based verification approach implemented in
CADP. Finally, Section~\ref{sec:concl} contains concluding remarks and outlines possible directions for
future work.

%
%
\section{Algorand Consensus Protocol}
\label{sec:background}
%
%

Algorand is a permissionless blockchain introduced in~\cite{CM19} that achieves consensus by using a pure
PoS mechanism. Its design aims to combine decentralization and security while maintaining performance even
with a high number of participants. Unlike traditional PoS systems, which rely on a fixed validator set or
require staking transactions to lock funds, Algorand uses a \emph{dynamic} and \emph{decentralized} method
to select small committees of participants in each round to propose and certify blocks. These committees
execute a Byzantine agreement protocol to reach consensus on a block, ensuring its commitment after a finite
number of steps, a property called \emph{finality}.

A key innovation in Algorand is the use of cryptographic \emph{self-sortition}, where all nodes in the
network independently check whether they have been selected to serve a specific role in the current round,
such as block proposer or committee verifier, without them being known to anyone else. This is done through
the use of VRFs, which allow each node to compute a pseudo-random, private, and uniquely verifiable
credential. The output of the VRF determines both whether the node is selected and the strength of its
selection (e.g., its weight as a verifier if holding multiple tokens). The credential serves as a proof of
selection and can be publicly verified by all the other participants.

This decentralized, private selection process achieves \emph{player replaceability} -- a core principle
emphasized in~\cite{CM19} -- which means that protocol security does not depend on the continuous
availability or honesty of specific, fixed participants. Since committee membership is privately determined
and changes at every step of the protocol, an adversary cannot know in advance the committee members,
therefore cannot corrupt or target them selectively. Once a node is selected, it reveals its credential
along with its vote; its influence on the consensus depends only on its stake, not on any persistent
identity.

Furthermore, the probability of being selected is proportional to the node's stake in the system, making the
protocol fair and secure against Sybil attacks~\cite{Dou02}, i.e., attacks where a malicious actor may join
the network with many public keys in order to steer the PoS consensus in its favor. The use of VRFs ensures
that each selection is unpredictable, locally computable, and globally verifiable, while keeping
communication overhead low. Cryptographic self-sortition underpins the block proposal phase and the
committee-based agreement that follows, forming the foundation of the Algorand approach to decentralized
consensus.

The protocol proceeds in rounds, each of which is related to a block and divided into two main phases: GC
and BBA$^{*}$~\cite{CM19}. The various steps of a round are executed by different, independently selected
committees, each chosen through the cryptographic self-sortition mechanism based on VRFs. The separation of
responsibilities between the GC and BBA$^{*}$ phases guarantees that honest verifiers never commit
conflicting blocks -- ensuring that Algorand never forks -- and that progress is always made either by
accepting a valid block or falling back to an empty one. This design also mitigates the risk posed by a
malicious or offline leader, because a faulty proposal cannot stall the protocol indefinitely.

In the remainder of this section, we present in more detail the GC (Section~\ref{sec:gc}) and BBA$^{*}$
(Section~\ref{sec:bba}) phases as they are implemented in Algorand and a brief discussion about the network
assumptions made in~\cite{CM19} (Section \ref{sec:algorand_assumptions}).

%
\subsection{GC -- Graded Consensus}
\label{sec:gc}
%

In the GC phase of round $r$, a leader $L^r$ is selected probabilistically and proposes a candidate block
$B^r$ by broadcasting it together with a valid VRF credential. A committee of verifiers, denoted by
$SV^{r,1}$ for step $1$ of round $r$, is selected to validate the proposal. These verifiers run a two-step
graded agreement protocol to attempt consensus on $B^r$. The outcome of this protocol is a \emph{grade} $g_i
\in \{ 0, 1, 2 \}$ computed individually by each verifier $i$ based on the messages it receives:

	\begin{itemize}

\item Grade $2$ indicates that the verifier has received strong support for the same proposed block from a
qualified majority of other verifiers. In particular, this corresponds to observing more than a
predetermined threshold $t$ of consistent votes for the block, which implies high confidence in global
agreement.

\item Grade $1$ indicates that the verifier has seen a majority vote, but not enough to meet the
high-confidence threshold $t$. The verifier recognizes some alignment, but not strong enough to safely
commit.

\item Grade $0$ means that the verifier has received highly inconsistent or insufficient information from
the other members of the committee, thus preventing any meaningful inference about consensus.

	\end{itemize}

Each verifier uses its grade output $g_i$ to determine its initial bit $b_i$ for the BBA$^{*}$ phase, which
is responsible for ensuring final agreement on $B^r$ or an empty block. Following the rule defined
in~\cite{CM19}, the initial bit is assigned as follows:
\[
b_i \: = \:
\begin{cases}
0 & \text{if } g_i = 2 \\
1 & \text{otherwise}
\end{cases}
\]
\noindent
A bit value of $0$ corresponds to confidence in the proposed block $B^r$, while $1$ indicates lack of
confidence, defaulting to support for the empty block.

%
\subsection{BBA$^{*}$ -- Binary Byzantine Agreement}
\label{sec:bba}
%

The BBA$^{*}$ protocol guarantees that all honest verifiers eventually reach agreement on a single binary
value, which determines whether to accept the candidate block $B^r$ proposed in the GC phase or commit
instead an empty block. This agreement is achieved under the assumptions that $(i)$ a large majority of
nodes are honest and $(ii)$ each verifier is selected independently and unpredictably via cryptographic
self-sortition. The protocol proceeds through a sequence of steps, each carried out by a newly and
independently selected committee of verifiers.

The BBA$^{*}$ protocol runs in cycles of three step types, called \emph{Coin-Fixed-To-0}, \linebreak
\emph{Coin-Fixed-To-1} and \emph{Coin-Genuinely-Flipped}. Each step type is carried out by a new committee
of verifiers $SV^{r,s}$ for round $r$ and step $s$. In every step, each node collects valid votes sent by
the committees of earlier steps and checks whether either \emph{ending condition 0} is met, meaning that the
node has received at least $t$ valid votes for bit $0$ from a prior \emph{Coin-Fixed-To-0} step, or
\emph{ending condition 1} \linebreak is met, meaning that the node has received at least $t$ valid votes for
bit $1$ from a prior \emph{Coin-Fixed-To-1} step. If either condition is satisfied, the node immediately
decides on the corresponding bit and halts, otherwise it updates its local bit according to the rules of the
current step type. The three step types alternate cyclically as follows:

	\begin{enumerate}

\item \emph{Coin-Fixed-To-0}: The node checks the two ending conditions. If neither holds it checks whether
it received at least $t$ valid votes for bit $1$ from the immediately preceding step, regardless of its
type. If this condition is met, the node sets its bit to $1$, otherwise it sets its bit to $0$, then it
proceeds to the next step. Following the notation of the corresponding table in~\cite[page $163$]{CM19}, a
step $s$ of this type corresponds to the class of steps for which $s - 2 \equiv 0 \mod 3$.

\item \emph{Coin-Fixed-To-1}: This step mirrors the previous one but inverts the bias. The node checks the
two ending conditions. If neither holds it checks whether it received at least $t$ valid votes for bit $0$
from the previous step. If so it sets its bit to $0$, otherwise it sets its bit to $1$, then it continues.
According to the aforementioned table in~\cite[page $163$]{CM19}, a step $s$ of this type corresponds to the
class of steps for which $s - 2 \equiv 1 \mod 3$.

\item \emph{Coin-Genuinely-Flipped}: In this step the node checks again both ending conditions. If neither
holds it then proceeds to update its bit to $0$ or $1$ by checking whether it has received at least $t$
valid votes for the corresponding bit in the previous step. If this further condition is not met, the bit is
chosen probabilistically. According to the aforementioned table in~\cite[page $163$]{CM19}, a step~$s$ of
this type corresponds to the class of steps for which $s - 2 \equiv 2 \mod 3$.

	\end{enumerate}

This cycle of step types is repeated until a final decision is made by all honest verifiers. Once consensus
is reached:

	\begin{itemize}

\item If the agreed bit is $0$, it means that verifiers trust the proposed block $B^r$ from the GC phase. In
this case, the block is accepted and a certificate is generated by collecting signatures from at least $t$
verifiers that voted to commit. This certificate proves finality and the block is appended to
the~blockchain.

\item If the agreed bit is $1$, it means that the candidate block was not supported with sufficient
confidence. Instead, verifiers commit a special empty block for round $r$, ensuring that the chain
progresses even in the presence of ambiguity or adversarial interference during the GC phase.

	\end{itemize}

%
\subsection{Network Assumptions and Their Consequences of Consensus}
\label{sec:algorand_assumptions}
%

In the original protocol, Algorand assumes a \emph{partially synchronous} network model~\cite{CM19}. It is
worth recalling that agreement can be reached in a synchronous network iff more than $2 / 3$ of the nodes
are working properly~\cite{PSL80,LSP82}, while it is impossible in an asynchronous network even with a
single faulty node~\cite{FLP85}. Similar results hold under partial synchrony~\cite{DLS88}
and randomization~\cite{Ben83,Rab83,Tou84}.

The considered model enables the protocol to operate correctly and
guarantee consensus in the presence of bounded message delays and asynchronous local clocks. These
assumptions are fundamental to the way Algorand structures its round-based consensus, allowing for the use
of cryptographic self-sortition, local decision making, and fault tolerance in a distributed setting. They
interact with the two main phases of the protocol as follows:

	\begin{itemize}

\item \emph{Bounded message delays:} The protocol assumes that control messages (e.g., graded consensus
results, sortition proofs, and votes) are delivered within a known time bound $\lambda$, while full block
proposals are delivered within a larger time bound $\Lambda$. These bounds are used to define local timeouts
for protocol transitions. Specifically:

		\begin{itemize}

\item In the GC phase, each node must wait up to $\Lambda$ to receive the block proposed by the leader and
its associated VRF credential before beginning validation.
 
\item In the BBA$^{*}$ phase, each node expects messages from other committee members within $\lambda$ to
determine whether they have received enough valid votes to satisfy either ending condition (i.e., at least
$t$ votes for a particular bit).
 
\item These time bounds ensure that honest nodes never wait indefinitely for a message. If a verifier does
not receive a message within the prescribed delay, it assumes that the message was not sent or that the
sender is faulty, then continues to the next step accordingly.

		\end{itemize}

\item \emph{Asynchronous clocks:} Algorand does not require nodes to maintain synchronized clocks.
Instead, each node proceeds through the protocol at its own pace, using local timers calibrated against
$\lambda$ and $\Lambda$. This enables several important behaviors:

		\begin{itemize}

\item The nodes independently determine when to start or end a given protocol step (e.g., step $s$ of round
$r$ of BBA$^{*}$) without coordinating with others.

\item The protocol remains robust against clock drift and execution delays, as each node progresses based on
the elapsed time since the last event and the expected delivery window for relevant messages.

\item The logical structure of the protocol -- organized in numbered steps inside rounds -- is maintained by
nodes based on local state and timeout checks, not global synchronization.

		\end{itemize}

	\end{itemize}

These assumptions have direct consequences on the internal logic of the Algorand consensus process. For
example, when a verifier at step $s$ of round $r$ of BBA$^{*}$ checks whether an ending condition is
satisfied, it does so by analyzing the messages received within a $\lambda$-bounded time window and
filtering only those messages that originate from the correct round $r$ and step $s-1$, contain valid VRF
credentials proving committee membership, and carry a bit value.

By relying on $\lambda$ and $\Lambda$ for timeout enforcement, Algorand ensures that steps progress
deterministically and safely. An adversary cannot exploit timing uncertainty to delay the protocol
indefinitely or to equivocate between subgroups of honest nodes. The bounded-delay model also supports the
use of player replaceability, because each committee in every step of every round is independently and
randomly selected, with no participant needing to persist across rounds or be globally coordinated. This
design achieves a high degree of decentralization, while remaining resilient against a powerful, adaptive
adversary capable of delaying -- but not indefinitely disrupting -- communication.

%
%
\section{Probabilistic Process Algebra Model of BBA$^{*}$}
\label{sec:model_algorand}
%
%

In this section we focus on BBA$^{*}$ and present our probabilistic process algebra model for it. The model
captures the essential structure and behavior of the protocol as used in Algorand, including the selection
of verifier committees, bit propagation, and step transitions. We consider a simplified case in which the
underlying network is fully synchronous and each stake unit is represented as an independent process whose
clock is synchronized with those of all the others. This abstraction allows us to model the consensus logic
in accordance with the theoretical results of~\cite{PSL80,LSP82,FLP85}, eliminating the need to account for
message delays or asynchronous progression. Our formalization faithfully reflects the alternation of the
three step types -- \emph{Coin-Fixed-To-0}, \emph{Coin-Fixed-To-1}, and \emph{Coin-Genuinely-Flipped} -- and
the rules governing the update and decision conditions for each verifier until either ending condition is
met.

We structure our study by first providing a formal model in the absence of malicious nodes, where all nodes
behave honestly and follow the protocol. This serves to validate the correct behavior and convergence of the
consensus mechanism under ideal conditions. We then extend the model to include an adversarial population
that controls at least $1 / 3$ of the total stake. Under this assumption, while still ensuring that no
conflicting blocks are ever committed, the protocol may fail to reach agreement on the proposed block. In
this case, the adversary can prevent consensus on bit~$0$ and bias the outcome toward bit~$1$, thus forcing
honest verifiers to commit an empty block. This adversarial influence can be interpreted as a form of
interference, where the adversary blocks progress and forces the commit on an empty block. This behavior
aligns with the fallback guarantees of the BBA$^{*}$ protocol and underscores the importance of maintaining
a supermajority of honest stake to preserve both safety and liveness, as originally formalized
in~\cite{CM19}.

In the remainder of this section, we first take a closer look at our assumptions and what they imply with
respect to the original protocol (Section~\ref{sec:model_assumptions}), then we introduce our probabilistic
process calculus (Section~\ref{sec:process_calculus}). Lastly we present the probabilistic process algebra
model by starting with a formalization where no malicious node is involved (Sections~\ref{sec:model_honest})
and then extending it with malicious nodes that can coordinate to force the commit of an empty block
(Section~\ref{sec:model_malicious}).

%
\subsection{Our Modeling Assumptions}
\label{sec:model_assumptions}
%

In the Algorand consensus protocol~\cite{CM19}, consensus security relies on the \emph{Honest Majority of
Money} (HMM): a supermajority (typically at least $2 / 3$) of the total stake must be controlled by honest
participants. This is implemented through cryptographic self-sortition, which selects committee members with
probability proportional to their stake. The protocol also implicitly assumes an \emph{Honest Majority of
Users} (HMU) by requiring that each user operates with a unique public key so as to avoid Sybil
attacks~\cite{Dou02}.

In our formal model, we approximate HMU by representing each minimal stake unit (or token) as an independent
user process. Each process independently runs self-sortition, sets its bit, and votes. This means that in
our model the total population size effectively corresponds to the total money supply, not the number of
distinct user identities. Consequently, our model does not explicitly enforce the HMU assumption. Instead,
we rely on the stronger HMM assumption, supposing that stake units (tokens) are independently controlled and
that the fraction of honest processes directly reflects the honest stake fraction.

To make the model tractable for formal verification, we also assume a \emph{fully synchronous} network
(instead of the partially synchronous one of Section~\ref{sec:algorand_assumptions}) in which:

	\begin{itemize}

\item All honest users (tokens) proceed synchronously at each step.

\item All messages are delivered instantly in each round.

\item There are no message delays, network faults, or asynchronous clocks.
  
	\end{itemize}

Under these stronger synchrony assumptions, the model simplifies several aspects of the protocol:

	\begin{itemize}

\item All users are always at the same step at the same time.
  
\item Ending conditions depend only on the messages received during the previous valid step.

\item No delayed or out-of-order messages occur, removing the need for explicit timeouts or recovery.

	\end{itemize}

%
\subsection{Our Probabilistic Process Calculus}
\label{sec:process_calculus}
%

In our formal model, we employ a probabilistic process calculus with standard operators taken
from~\cite{Mil89a,BHR84,HJ90,AGT12}, which includes an action prefix operator, a nondeterministic choice
operator, a probabilistic choice operator, and a CSP-style parallel composition operator. 

Let $\cala$ be the set of observable action labels. We define the extended action set $\cala_{\tau}$ by
adding the distinguished unobservable action $\tau$, so that $\cala_{\tau} = \cala \cup \{ \tau \}$. The set
$\procs$ of process terms is generated by the following grammar:
\cws{0}{P \: ::= \: \nil \mid \beta \rightarrow P \mid a \, . \, P \mid P + P \mid [p] P \oplus [1 - p] P
\mid P \pco{L} P \mid P \setminus L \mid P \, / \, L}
where:

	\begin{itemize}

\item $\nil$ is the terminated process.

\item $\beta \rightarrow P$ behaves as process $P$ if the boolean guard $\beta$ is true.

\item $a \, . \, \_$, for $a \in \cala_{\tau}$, is the action prefix operator describing a process that can
initially perform action $a$.

\item $\_ + \_$ is the alternative composition operator expressing a choice between two processes. We write
$\sum\limits_{k = 1}^{n} P_{i}$ as a shorthand for $P_{1} + P_{2} + \dots + P_{n}$.

\item $[p] \, \_ \oplus [1 - p] \, \_$ is the probabilistic composition operator expressing a probabilistic
choice between two processes with probabilities $p, 1 - p \in \realns_{]0, 1[}$ summing up to $1$.

\item $\_ \pco{L} \_$, for $L \subseteq \cala$, is the parallel composition operator allowing two processes
to proceed independently on any action not in $L$ and forcing them to synchronize on every action in $L$.
We write $\pco{L}\limits_{\hspace*{-5pt}i = 1}^{\hspace*{-5pt}n} P_{i}$ as a shorthand for $P_{1} \pco{L}
P_{2} \pco{L} \dots \pco{L} P_{n}$.

\item $\_ \setminus L$, for $L \subseteq \cala$, is the restriction operator, which prevents the execution
of all actions belonging to $L$.

\item $\_ \, / \, L$, for $L \subseteq \cala$, is the hiding operator, which turns all the executed actions
belonging to $L$ into the unobservable action $\tau$. 

	\end{itemize}

The hiding and restriction operators are necessary to express noninterference properties, as we will see in
Section~\ref{sec:cadp}.

%
\subsection{BBA$^{*}$ Model without Malicious Nodes}
\label{sec:model_honest}
%

The consensus protocol is modeled as the parallel composition of several processes, each representing a
distinct component of the system:

	\begin{itemize}

\item Process $\mathit{Node}_{i}$ represents an individual stake unit (token) acting as an independent
participant in consensus. $\mathit{Node}_{i}$ executes local actions such as receiving block proposals,
computing its initial bit, verifying committee membership, propagating its vote, and participating in the
binary agreement steps until either ending condition is met.
  
\item Processes $N_{i}, N'_{i}, N''_{i, s, b}, N'''_{i, s}, N''''_{i, s}$ define the internal states of the
sequential behavior of $\mathit{Node}_{i}$. These processes describe how node $i$ computes its bit $b$,
adjusts it according to the type of step $s$, and verifies conditions for committing the proposed or empty
block.
  
\item Process $C_{i, k_{0}, k_{1}}$ is a counter that keeps track for $\mathit{Node}_{i}$ of the number of
received votes for bit $0$ and bit $1$. This allows $\mathit{Node}_{i}$ to verify whether the threshold $t$
for committing a decision has been reached or not.

\item Action $\mathit{self\_verify}$ models self-sortition for $\mathit{Node}_{i}$, with $p_{\mathit{in}}$
and $p_{\mathit{out}}$ being the probabilities of belonging or not to the committee for the current step,
followed by $C_{i, k_{0}, k_{1}}$ reset.

\item Action $\mathit{propagate}_{i, b}$ models the broadcast of bit $b$ of $\mathit{Node}_{i}$ to the other
nodes.
  
\item Actions $\mathit{ask}_{0}, \mathit{ask}_{1}, \mathit{reply}_{h}$ represent local request and response
actions for counting how many bits have been received in the previous step.
  
\item Action $\mathit{sync}$ is a synchronization action ensuring that all nodes proceed together from a
protocol step to the next one under the fully synchronous execution assumption. Its two occurrences before
and after bit propagation ensure a correct bit counting.

\item Actions $\mathit{compute\_bit}$ and $\mathit{adjust\_bit}$ respectively model the initial bit computed
in the GC phase and its possible subsequent adjustments, where $p_{0}$ and $p_{1}$ represent the
probabilities that $\mathit{Node}_{i}$ initially sets its bit to $0$ or $1$. 
  
\item Actions $\mathit{receive\_block\_proposal}$, $\mathit{commit\_proposed\_block}$,
$\mathit{commit\_empty\_block}$ represent the initial and final actions of the current round. If the nodes
reach consensus on bit $0$, they commit the proposed block; if consensus is instead reached on bit $1$, they
commit an empty block.

	\end{itemize}

Together, these processes describe how stake units interact, how decisions propagate through the network,
and how consensus emerges under the assumed synchronous conditions. What follows is the general structure of
the protocol:
\[
\begin{array}{rcl}
\mathit{Algorand} & \: \eqdef \: & \pco{S}\limits_{\hspace*{-5pt}i=1}^{\hspace*{-5pt}n} \mathit{Node}_{i} \\

& &	\begin{array}{rcl}

\text{ where } S & = & \{ \mathit{receive\_block\_proposal}, \mathit{sync} \} \, \cup \\
& & \{ \mathit{propagate}_{i, 0}, \mathit{propagate}_{i, 1} \mid 1 \leq i \leq n \} \, \cup \\
& & \{ \mathit{commit\_proposed\_block}, \mathit{commit\_empty\_block} \} \\

	\end{array} \\

\mathit{Node}_{i} & \: \eqdef \: & N_{i} \pco{S'} C_{i, 0, 0} \\

& &  	\begin{array}{rcl}

\text{ where } S' & = & \{ \mathit{propagate}_{i, 0}, \mathit{propagate}_{i, 1}, \mathit{ask}_{0},
\mathit{ask}_{1} \} \, \cup \\
& & \{ \mathit{reply}_{j} \mid 0 \leq j \leq n \} \cup \{ \mathit{self\_verify} \} \\

	\end{array} \\

\end{array}
\]

\noindent
The next behavioral equations define the internal states that each node traverses during the execution of
the protocol ($t = \frac{2}{3} \cdot c$ where $c$ is the committee size, \linebreak the bit value $0$ or $1$
is in boldface when it is computed for the first time or adjusted afterwards):
\[
\begin{array}{lclr}
N_{i} & \eqdef \, & \mathit{receive\_block\_proposal} \, . \, N'_{i} & \\

N'_{i} & \eqdef \, & \mathit{compute\_bit} \, . \, ([p_{0}] N''_{i, \mathbf{0}} \oplus [p_{1}] N''_{i,
\mathbf{1}}) & \\

N''_{i, b} & \eqdef \, & \mathit{self\_verify} \, . \, \mathit{sync} \, . \, ([p_{\mathit{in}}]
\mathit{propagate}_{i, b} \, . \, \mathit{sync} \, . \, N'''_{i, \equiv 0} \oplus [p_{\mathit{out}}]
\mathit{sync} \, . \, N'''_{i, \equiv 0}) \\

N'''_{i, \equiv 0} & \eqdef \, & \mathit{ask}_{0} \, . \, \sum\limits_{k = 0}^{n} \mathit{reply}_{k} \, . \,
(k \geq t \rightarrow \mathit{commit\_proposed\_block} \, . \, N_{i} \, + \\
& & \hspace*{80pt} k < t \rightarrow N''''_{i, \equiv 0}) \\

N''''_{i, \equiv 0} & \eqdef \, & \mathit{adjust\_bit} \, . \, \mathit{ask}_{1} \, . \, \sum\limits_{k =
0}^{n} \mathit{reply}_{k} \, . \, (k \geq t \rightarrow N''_{i, \equiv 0, \mathbf{1}} \, + \\
& & \hspace*{129pt} k < t \rightarrow N''_{i, \equiv 0, \mathbf{0}}) \\

N''_{i, \equiv 0, b} & \eqdef \, & \mathit{self\_verify} \, . \, \mathit{sync} \, . \, ([p_{\mathit{in}}]
\mathit{propagate}_{i, b} \, . \, \mathit{sync} \, . \, N'''_{i, \equiv 1} \oplus [p_{\mathit{out}}]
\mathit{sync} \, . \, N'''_{i, \equiv 1}) \\

N'''_{i, \equiv 1} & \eqdef \, & \mathit{ask}_{1} \, . \, \sum\limits_{k = 0}^{n} \mathit{reply}_{k} \, . \,
(k \geq t \rightarrow \mathit{commit\_empty\_block} \, . \, N_{i} \, + \\
& & \hspace*{80pt} k < t \rightarrow N''''_{i, \equiv 1}) \\

N''''_{i, \equiv 1} & \eqdef \, & \mathit{adjust\_bit} \, . \, \mathit{ask}_{0} \, . \, \sum\limits_{k =
0}^{n} \mathit{reply}_{k} \, . \, (k \geq t \rightarrow N''_{i, \equiv 1, \mathbf{0}} \, + \\
& & \hspace*{129pt} k < t \rightarrow N''_{i, \equiv 1, \mathbf{1}}) \\

N''_{i, \equiv 1, b} & \eqdef \, & \mathit{self\_verify} \, . \, \mathit{sync} \, . \, ([p_{\mathit{in}}]
\mathit{propagate}_{i, b} \, . \, \mathit{sync} \, . \, N''''_{i, \equiv 2} \oplus [p_{\mathit{out}}]
\mathit{sync} \, . \, N''''_{i, \equiv 2}) \\

N''''_{i, \equiv 2} & \eqdef \, & \mathit{adjust\_bit} \, . \, (\mathit{ask}_{0} \, . \, \sum\limits_{k =
0}^{n} \mathit{reply}_{k} \, . \, (k \geq t \rightarrow N''_{i, \mathbf{0}} \, + \\
& & \hspace*{119pt} k < t \rightarrow \mathit{ask}_{1} \, . \, \sum\limits_{k = 0}^{n} \mathit{reply}_{k} \,
. \, (k \geq t \rightarrow N''_{i, \mathbf{1}} \, + \\
& & \hspace*{237pt} k < t \rightarrow N'_{i}))) \\
\end{array}
\]

\noindent	
Lastly, we have the definition of the counter, which keeps track for the corresponding node of the number of
received bits equal to $0$ or $1$:

\[
\begin{array}{lcl}	
C_{i, k_{0}, k_{1}} & \eqdef & \sum\limits_{j \in \{ 1, \dots, n \}} \mathit{propagate}_{j, 0} \, . \, C_{i,
k_{0} + 1, k_{1}} \\
& + & \sum\limits_{j \in \{ 1, \dots, n \}} \mathit{propagate}_{j, 1} \, . \, C_{i, k_{0}, k_{1} + 1} \\
& + & \mathit{ask}_{0} \, . \, \mathit{reply}_{k_{0}} \, . \, C_{i, k_{0}, k_{1}} \\
& + & \mathit{ask}_{1} \, . \, \mathit{reply}_{k_{1}} \, . \, C_{i, k_{0}, k_{1}} \\
& + & \mathit{self\_verify} \, . \, C_{i, 0, 0} \\
\end{array}
\]

%
\subsection{BBA$^{*}$ Model with Malicious Nodes}
\label{sec:model_malicious}
%

We now extend our formalization to include the behavior of malicious nodes. These adversarial participants
can inspect the proposed block and coordinate a boycott to prevent its commitment, thereby forcing the
commit of an empty block instead. More precisely:

	\begin{itemize}

\item $\mathit{MNode}_{i}, \mathit{MN}_{i}, \mathit{HN}_{i}$ are additional processes used in the extended
model with malicious behavior. $\mathit{MNode}_{i}$ is a malicious node that can boycott or deviate from
honest behavior, while $\mathit{HN}_{i}$ and $\mathit{MN}_{i}$ encode honest and malicious variants of the
internal states of node $i$.

\item $\mathit{boycott}$ is special action used to model the coordination of the malicious nodes, relevant
for non-interference and security analysis.

	\end{itemize}

\noindent
The extended model is as follows:

\[
\begin{array}{rcl}

\mathit{MAlgorand} & \: \eqdef \: & (\pco{S_{1 .. n}}\limits_{\hspace*{-20pt}i = 1}^{\hspace*{-20pt}n}
\mathit{Node}_{i}) \pco{S_{1 .. n + m}} (\pco{S_{n + 1 .. n + m} \cup \{ \mathit{boycott}
\}}\limits_{\hspace*{-78.3pt}i = n + 1}^{\hspace*{-78.3pt}n + m} \mathit{MNode}_{i})  \\
	
& & 	\begin{array}{rcl}
 
\text{ where } S_{l .. l'} & = & \{ \mathit{receive\_block\_proposal}, \mathit{sync} \} \, \cup \\
& & \{ \mathit{propagate}_{i, 0}, \mathit{propagate}_{i, 1} \mid l \leq i \leq l' \} \, \cup \\
& & \{ \mathit{commit\_proposed\_block}, \mathit{commit\_empty\_block} \} \\

	\end{array} \\

\mathit{MNode}_{i} & \eqdef & \mathit{MN}_{i, 0} \pco{S'_{n + m}} C_{i, 0, 0} \\

& & 	\begin{array}{rcl}
 
\text{ where } S'_{l} & = & \{ \mathit{propagate}_{i, 0}, \mathit{propagate}_{i, 1}, \mathit{ask}_{0},
\mathit{ask}_{1} \} \, \cup \\
& & \{ \mathit{reply}_{j} \mid 0 \leq j \leq l \} \cup \{ \mathit{self\_verify} \} \\

	\end{array} \\

\mathit{MN}_{i} & \eqdef & \mathit{receive\_block\_proposal} \, . \, (\mathit{boycott} \, . \, 
\mathit{MN}'_{i} + \tau \, . \, \mathit{HN}'_{i}) \\

\end{array}
\]

Each $\mathit{MN}_{i, s}$ is obtained from $\mathit{N}_{i, s}$ by replacing every occurrence of
$\mathit{propagate}_{i, b}$ with $\mathit{propagate}_{i, 1}$ and every instance of $N$ with $\mathit{MN}$.
Each $\mathit{HN}_{i, s}$ is similarly derived from $\mathit{N}_{i, s}$, but by replacing every occurrence
of $N$ with $\mathit{HN}$, except for $N_{i, 0}$ that becomes $\mathit{MN}_{i, 0}$. In both cases, every
occurrence of $n$ is replaced by $n + m$. The counter $C_{i, k_{0}, k_{1}}$ remains unchanged, except that
all occurrences of~$n$ are replaced with $n + m$ to reflect the total number of nodes.

%
%
\section{Verification}
\label{sec:verification}
%
%

In this section we present the translation of our probabilistic process algebra model into the specification
language LNT of CADP, along with a script to verify the properties of interest. We consider a very simple
case, with a network comprising only four nodes. Even with such a small number of nodes, it can be shown
that if there is more than one malicious node the consensus protocol can be hindered and the proposal of an
empty block can be forced.

Our LNT translation follows very closely the model given in the Sections~\ref{sec:model_honest}
and~\ref{sec:model_malicious}. We first show the translation of honest nodes under the ideal behavior of the
protocol, then we exhibit how to obtain the translation of malicious nodes by modifying the honest ones. For
each LNT process, we show the correspondence with its pure process algebraic counterpart.

In the remainder of this section, we first explain our choices regarding the parameters used in our model
(Section~\ref{sec:model_parameters}), then we show our LNT translation of both honest and malicious nodes
(Sections~\ref{sec:lnt} and~\ref{sec:lnt_mal}), and finally we provide a script to check noninterference
properties on our LNT translation in order to verify under which circumstances the malicious nodes can
hinder the block proposal (Section~\ref{sec:cadp}).

%
\subsection{Model Parameters}
\label{sec:model_parameters}
%

In our model, we assume a total population of $n = 4$ public keys in the network, representing the number of
stake-holding nodes or tokens participating in consensus. This does not reflect a realistic deployment
scenario for a large-scale permissionless blockchain such as Algorand, where the number of accounts with
meaningful stake can reach into the millions, but it is a big enough size to show how a group of malicious
nodes can boycott the commit of a proposed block.

We target an expected committee size of $c = 3$ verifiers per step. This committee size represents a
tradeoff between security and efficiency: it is large enough to ensure a high probability of an honest
majority, while still small enough to keep communication and verification overhead manageable.

Given these values, the probability that any single node (or unit of stake) \linebreak is selected to
participate in the committee of a given step as a verifier is:
\[
p_{v} = \frac{c}{n} = 0.75
\]
This probability is used in the model to determine whether a given process takes part in a specific step in
the role of a verifier. Since cryptographic self-sortition is executed independently by each node, this
selection is modeled as a local probabilistic choice by each process.

In addition, consistent with the security assumptions of~\cite{CM19}, which guarantee safety and liveness
provided that the majority of money is controlled by honest nodes, we expect the \emph{honest money
fraction} to be $h = 0.8$, meaning that $80\%$ of the total stake should belong to honest participants;
notice that this estimation gives no guarantee or bound about the actual number of honest/malicious nodes
involved in the protocol. We then compute the probability $p_{h}$ that a selected leader is honest by using
the following formula given in~\cite{CM19}:
\[
p_{h} = h^{2} \cdot (1 + h - h^{2}) = 0.7424
\]
This value plays an indirect role in the bit-setting logic used in the BBA$^{*}$ phase. In particular, the
probability that a verifier sets its initial bit to $0$ (i.e., supports the proposed block) depends on
whether the block proposer in the GC phase was honest and whether enough honest verifiers supported the
proposal. As described in~\cite[Section~$4$]{CM19}, an initial bit of $0$ corresponds to receiving a grade
$g_{i} = 2$, \linebreak which occurs when the GC committee includes an honest leader and a strong honest
majority of votes. The resulting bit-setting behavior is thus captured probabilistically in our model by
employing $p_{h}$ as a proxy for a successful GC~phase.

The parameters $p_{v}$ for committee selection and $p_{h}$ for initial bit assignment are used to guide all
probabilistic choices in our model.

%
\subsection{LNT Translation}
\label{sec:lnt}
%

Our LNT model consists of two main files. The first file, named \texttt{DATA.lnt}, comprises channel
declarations for the various actions appearing in the processes, along with \texttt{N} and \texttt{T}
respectively describing the constants for the maximum number of nodes in the network and the vote threshold
needed to commit a block. In this instance, such values are set to $4$ and $2$ respectively:

	{\small\begin{verbatim}
  module DATA (BIT) is

     -- The BIT type is imported from the predefined BIT library

     -- Type declaration for the various kinds of steps of BBA*
     type STEP is
        S_INIT,   -- initialization step
        S_ZERO,   -- congruent-to-0-modulo-3 step
        S_ONE,    -- congruent-to-1-modulo-3 step
        S_TWO     -- congruent-to-2-modulo-3 step
        with =, <>
     end type

     -- Channel declaration for block proposal or commit actions
     channel BLOCK is
        ()
     end channel

     -- Channel declaration for synchronization actions
     channel SYNCHRONIZE is
        ()
     end channel


     -- Channel declaration for bit computation actions
     channel COMPUTE is
        ()
     end channel
	\end{verbatim}}

	{\small\begin{verbatim}
     -- Channel declaration for bit propagation actions
     -- ID is the node identifier and B is the propagated bit
     channel PROPAGATE is
        (ID: NAT, B: BIT)
     end channel
	\end{verbatim}}

	{\small\begin{verbatim}
     -- Channel declaration for bit self-propagation actions
     -- (self-propagation goes from each node to its counter)
     channel SELF_PROPAGATE is
        (B: BIT)
     end channel
	\end{verbatim}}

	{\small\begin{verbatim}
     -- Channel declaration for ask actions
     channel ASK is
        (B: BIT)
     end channel
	\end{verbatim}}

	{\small\begin{verbatim}
     -- Channel declaration for reply actions
     channel REPLY is
        (N: NAT)
     end channel
	\end{verbatim}}

	{\small\begin{verbatim}
     -- Channel declaration for self-verification actions
     channel VERIFY is
        ()
     end channel
	\end{verbatim}}

	{\small\begin{verbatim}
     -- Channel declaration for probabilistic choices
     channel PROBABILISTIC is
        (P: REAL)
     end channel
	\end{verbatim}}

	{\small\begin{verbatim}
     -- Channel declaration for boycott actions
     channel BOYCOTT is
        ()
     end channel
	\end{verbatim}}

	{\small\begin{verbatim}
     -- Maximum number of nodes in the network
     function N: NAT is
        return 4
     end function
	\end{verbatim}}

	{\small\begin{verbatim}
     -- Threshold of votes needed to commit a block
     function T: NAT is
        return 2
     end function
	\end{verbatim}}

	{\small\begin{verbatim}
  end module
	\end{verbatim}}

\pagebreak

The second file, named \texttt{ALGORAND.lnt}, comprises the translations for the various processes of our
probabilistic process algebra model. We start by showing the translation of the process $\mathit{Node}_{i}$
of Section~\ref{sec:model_honest}. This process is given by the parallel composition of a subprocess
\texttt{N}, which is its behavioral component, and a subprocess \texttt{C}, which is its counter. Each node
has an argument \texttt{ID}, which is a natural number representing the identifier of the node:

	{\small\begin{verbatim}
  process NODE [RECEIVE_BLOCK_PROPOSAL, COMMIT_PROPOSED_BLOCK,
                COMMIT_EMPTY_BLOCK: BLOCK, SYNC: SYNCHRONIZE,
                COMPUTE_BIT, ADJUST_BIT: COMPUTE,
                PROPAGATE: PROPAGATE, SELF_PROPAGATE: SELF_PROPAGATE,
                ASK: ASK, REPLY: REPLY, SELF_VERIFY: VERIFY,
                P_B, P_IN, P_OUT: PROBABILISTIC]
               (ID: NAT) is

     par SELF_PROPAGATE, ASK, REPLY, SELF_VERIFY in
        N [...] (ID)
     ||
        C [...] (ID, 0, 0)
     end par

  end process
	\end{verbatim}}

The next two processes model the receiving of the block proposal and the initial computation of the bit by
each node. They implement processes $N_{i}$ and~$N'_{i}$ of Section~\ref{sec:model_honest}:

	{\small\begin{verbatim}
  process N [RECEIVE_BLOCK_PROPOSAL, COMMIT_PROPOSED_BLOCK,
             COMMIT_EMPTY_BLOCK: BLOCK, SYNC: SYNCHRONIZE,
             COMPUTE_BIT, ADJUST_BIT: COMPUTE,
             PROPAGATE: PROPAGATE, SELF_PROPAGATE: SELF_PROPAGATE,
             ASK: ASK, REPLY: REPLY, SELF_VERIFY: VERIFY,
             P_B, P_IN, P_OUT: PROBABILISTIC]
            (ID: NAT) is

     RECEIVE_BLOCK_PROPOSAL;
     N_PRIME [...] (ID)

  end process

  process N_PRIME [RECEIVE_BLOCK_PROPOSAL, COMMIT_PROPOSED_BLOCK,
                   COMMIT_EMPTY_BLOCK: BLOCK, SYNC: SYNCHRONIZE,
                   COMPUTE_BIT, ADJUST_BIT: COMPUTE,
                   PROPAGATE: PROPAGATE, SELF_PROPAGATE: SELF_PROPAGATE,
                   ASK: ASK, REPLY: REPLY, SELF_VERIFY: VERIFY,
                   P_B, P_IN, P_OUT: PROBABILISTIC]
                  (ID: NAT) is

     COMPUTE_BIT;
     var P_0: REAL in
        P_0 := 0.7424;
        alt
           P_B (P_0);
           N_SECOND [...] (ID, S_INIT, 0)
        []
           P_B (1.0 - P_0);
           N_SECOND [...] (ID, S_INIT, 1)
        end alt
     end var

  end process
	\end{verbatim}}

The next process models the cryptographic self-sortition that each node performs to check whether it is a
committee member for that step or not. If a node is selected to be in the committee, it then propagates its
bit along with its identifier, otherwise it simply proceeds forward to the next process. This translates
processes $N''_{i, b}, N''_{i, \equiv 0, b}, N''_{i, \equiv 1, b}$ of Section~\ref{sec:model_honest}. Along
with the identifier \texttt{ID}, two additional arguments are required, respectively called \texttt{S},
representing the step the node is currently in, and \texttt{B}, representing the bit it is going to
propagate:

	{\small\begin{verbatim}
  process N_SECOND [RECEIVE_BLOCK_PROPOSAL, COMMIT_PROPOSED_BLOCK,
                    COMMIT_EMPTY_BLOCK: BLOCK, SYNC: SYNCHRONIZE,
                    COMPUTE_BIT, ADJUST_BIT: COMPUTE,
                    PROPAGATE: PROPAGATE, SELF_PROPAGATE: SELF_PROPAGATE,
                    ASK: ASK, REPLY: REPLY, SELF_VERIFY: VERIFY,
                    P_B, P_IN, P_OUT: PROBABILISTIC]
                   (ID: NAT, S: STEP, B: BIT) is

     require S <> S_TWO;

     var IN: REAL in
        IN := 0.75;
        SELF_VERIFY;
        SYNC;
        alt
           P_IN (IN);
           PROPAGATE (ID, B);
           SELF_PROPAGATE (B)
        []
           P_OUT (1.0 - IN)
        end alt;
        SYNC;
        case S in
           S_INIT ->
              N_THIRD [...] (ID, S_ZERO)
        |  S_ZERO ->
              N_THIRD [...] (ID, S_ONE)
        |  S_ONE ->
              N_FOURTH [...] (ID, S_TWO)

        |  any ->
              raise UNEXPECTED
        end case
     end var

  end process
	\end{verbatim}}

The following process models the interaction between a node and its counter and the subsequent commit of a
block. It translates processes $N'''_{i, \equiv 0}$ and $N'''_{i, \equiv 1}$ of
Section~\ref{sec:model_honest}:

	{\small\begin{verbatim}
  process N_THIRD [RECEIVE_BLOCK_PROPOSAL, COMMIT_PROPOSED_BLOCK,
                   COMMIT_EMPTY_BLOCK: BLOCK, SYNC: SYNCHRONIZE,
                   COMPUTE_BIT, ADJUST_BIT: COMPUTE,
                   PROPAGATE: PROPAGATE, SELF_PROPAGATE: SELF_PROPAGATE,
                   ASK: ASK, REPLY: REPLY, SELF_VERIFY: VERIFY,
                   P_B, P_IN, P_OUT: PROBABILISTIC]
                  (ID: NAT, S: STEP) is

     require S = S_ZERO or S = S_ONE;

     var K: NAT in
        case S in
           S_ZERO ->
              ASK (0 of BIT);
              REPLY (?K) where K <= N;
              if K >= T then
                 COMMIT_PROPOSED_BLOCK;
                 N [...] (ID)
              else
                 N_FOURTH [...] (ID, S_ZERO)
              end if
        |  S_ONE ->
              ASK (1 of BIT);
              REPLY (?K) where K <= N;
              if K >= T then
                 COMMIT_EMPTY_BLOCK;
                 N [...] (ID)
              else
                 N_FOURTH [...] (ID, S_ONE)
              end if
        |  any ->
              raise UNEXPECTED
        end case
     end var

  end process
	\end{verbatim}}

The following process models the adjustment of the bit that each node performs if a commit failed. It
translates processes $N''''_{i, \equiv 0}, N''''_{i, \equiv 1}, N''''_{i, \equiv 2}$ of
Section~\ref{sec:model_honest}:

	{\small\begin{verbatim}
  process N_FOURTH [RECEIVE_BLOCK_PROPOSAL, COMMIT_PROPOSED_BLOCK,
                    COMMIT_EMPTY_BLOCK: BLOCK, SYNC: SYNCHRONIZE,
                    COMPUTE_BIT, ADJUST_BIT: COMPUTE,
                    PROPAGATE: PROPAGATE, SELF_PROPAGATE: SELF_PROPAGATE,
                    ASK: ASK, REPLY: REPLY, SELF_VERIFY: VERIFY,
                    P_B, P_IN, P_OUT: PROBABILISTIC]
                   (ID: NAT, S: STEP) is

     require S <> S_INIT;

     ADJUST_BIT;
     var K: NAT in
        case S in
           S_ZERO ->
              ASK (1 of BIT);
              REPLY (?K) where K <= N;
              if K >= T then
                 N_SECOND [...] (ID, S_ZERO, 1)
              else
                 N_SECOND [...] (ID, S_ZERO, 0)
              end if
        |  S_ONE ->
              ASK (0 of BIT);
              REPLY (?K) where K <= N;
              if K >= T then
                 N_SECOND [...] (ID, S_ONE, 0)
              else
                 N_SECOND [...] (ID, S_ONE, 1)
              end if
        |  S_TWO ->
              ASK (0 of BIT);
              REPLY (?K) where K <= N;
              if K >= T then
                 N_SECOND [...] (ID, S_INIT, 0)
              else
                 ASK (1 of BIT);
                 REPLY (?K) where K <= N;
                 if K >= T then
                    N_SECOND [...] (ID, S_INIT, 1)
                 else
                    N_PRIME [...] (ID)
                 end if
              end if
        |  any ->
              raise UNEXPECTED
        end case
     end var

  end process
	\end{verbatim}}

Lastly, we have the counter that interacts with the behavioral part of the corresponding node to keep track
of the bits propagated by the other nodes. \linebreak It translates process $C$ of
Section~\ref{sec:model_honest}:

	{\small\begin{verbatim}
  process C [PROPAGATE: PROPAGATE, SELF_PROPAGATE: SELF_PROPAGATE,
             ASK: ASK, REPLY: REPLY, SELF_VERIFY: VERIFY]
            (ID, K_0, K_1: NAT) is

     var J: NAT, B: BIT in
        alt
           alt
              PROPAGATE (?J, ?B) where 1 <= J and J <= N and J <> ID
           []
              SELF_PROPAGATE (?B)
           end alt;
           if B == 0 then
              C [...] (ID, K_0 + 1, K_1)
           else
              C [...] (ID, K_0, K_1 + 1)
           end if
        []
           ASK (?B);
           if B == 0 then
              REPLY (K_0)
           else
              REPLY (K_1)
           end if;
           C [...] (ID, K_0, K_1)
        []
           SELF_VERIFY;
           C [...] (ID, 0, 0)
        end alt
     end var

  end process
	\end{verbatim}}

To represent the behavior of the whole protocol, and hence to implement the process $\mathit{Algorand}$ of
Section~\ref{sec:model_honest}, we define the main process given by a parallel composition of a certain
number of nodes, which is $4$ in our case:

	{\small\begin{verbatim}
  process MAIN [RECEIVE_BLOCK_PROPOSAL, COMMIT_PROPOSED_BLOCK, 
                COMMIT_EMPTY_BLOCK: BLOCK, SYNC: SYNCHRONIZE,
                COMPUTE_BIT, ADJUST_BIT: COMPUTE,
                PROPAGATE: PROPAGATE, SELF_PROPAGATE: SELF_PROPAGATE,
                ASK: ASK, REPLY: REPLY, SELF_VERIFY: VERIFY,
                P_B, P_IN, P_OUT: PROBABILISTIC] is

     par RECEIVE_BLOCK_PROPOSAL, PROPAGATE, SYNC,
         COMMIT_PROPOSED_BLOCK, COMMIT_EMPTY_BLOCK in

        NODE [...] (1)
     ||
        NODE [...] (2)
     ||
        NODE [...] (3)
     ||
        NODE [...] (4)
     end par

  end process
	\end{verbatim}}

%
\subsection{LNT Translation of Malicious Nodes}
\label{sec:lnt_mal}
%

Starting from the previous LNT specifications, we can translate the behavior of a malicious node as defined
in Section~\ref{sec:model_malicious}. Similar to what we have done with \texttt{NODE} and \texttt{N}, we can
implement their malicious counterparts \texttt{MNODE} and \texttt{MN}:

	{\small\begin{verbatim}
  process MNODE [RECEIVE_BLOCK_PROPOSAL, COMMIT_PROPOSED_BLOCK,
                 COMMIT_EMPTY_BLOCK: BLOCK, SYNC: SYNCHRONIZE,
                 COMPUTE_BIT, ADJUST_BIT: COMPUTE,
                 PROPAGATE: PROPAGATE, SELF_PROPAGATE: SELF_PROPAGATE,
                 ASK: ASK, REPLY: REPLY, SELF_VERIFY: VERIFY,
                 P_B, P_IN, P_OUT: PROBABILISTIC, BOYCOTT: BOYCOTT]
                (ID: NAT) is

     par SELF_PROPAGATE, ASK, REPLY, SELF_VERIFY in
        MN [...] (ID)
     ||
        C [...] (ID, 0, 0)
     end par

  end process
	\end{verbatim}}

In process \texttt{MN} there is a choice between the action \texttt{BOYCOTT} and the internal action
\texttt{i}, representing the behavior of malicious nodes that can decide or not to coordinate in order to
hinder the commit of the proposed block. We remind that for noninterference purposes we will regard
\texttt{BOYCOTT} as a high-level action:

	{\small\begin{verbatim}
  process MN [RECEIVE_BLOCK_PROPOSAL, COMMIT_PROPOSED_BLOCK,
              COMMIT_EMPTY_BLOCK: BLOCK, SYNC: SYNCHRONIZE,
              COMPUTE_BIT, ADJUST_BIT: COMPUTE,
              PROPAGATE: PROPAGATE, SELF_PROPAGATE: SELF_PROPAGATE,
              ASK: ASK, REPLY: REPLY, SELF_VERIFY: VERIFY,
              P_B, P_IN, P_OUT: PROBABILISTIC, BOYCOTT: BOYCOTT]
             (ID: NAT) is

     RECEIVE_BLOCK_PROPOSAL;
     alt

        BOYCOTT;
        MN_PRIME [...] (ID)
     []
        i;
        HN_PRIME [...] (ID)
     end alt

  end process
	\end{verbatim}}

\texttt{MN\_PRIME} and the subsequent \texttt{MN\_SECOND}, \texttt{MN\_THIRD}, \texttt{MN\_FOURTH} are
obtained from their honest counterparts by replacing each occurrence of \texttt{B} with $1$. The process
\texttt{HN\_PRIME} and the subsequent \texttt{HN\_SECOND}, \texttt{HN\_THIRD}, \texttt{HN\_FOURTH},
representing the case in which the malicious nodes do not coordinate to boycott the commit of the block, are
identical to \texttt{N\_PRIME} and the subsequent \texttt{N\_SECOND}, \texttt{N\_THIRD}, \texttt{N\_FOURTH}
with the only difference being that \texttt{N} is replaced by \texttt{MN}.

Finally, we translate process $\mathit{MAlgorand}$ of Section~\ref{sec:model_malicious}, representing the
behavior of the protocol with both honest and malicious nodes. In the new file \texttt{MALGORAND.lnt} we
consider again the case in which the total size of the network is $4$, with $2$~honest nodes and
$2$~malicious nodes:

	{\small\begin{verbatim}
  process MAIN [RECEIVE_BLOCK_PROPOSAL, COMMIT_PROPOSED_BLOCK,
                COMMIT_EMPTY_BLOCK: BLOCK, SYNC: SYNCHRONIZE,
                COMPUTE_BIT, ADJUST_BIT: COMPUTE,
                PROPAGATE: PROPAGATE, SELF_PROPAGATE: SELF_PROPAGATE,
                ASK: ASK, REPLY: REPLY, SELF_VERIFY: VERIFY,
                P_B, P_IN, P_OUT: PROBABILISTIC, BOYCOTT: BOYCOTT] is

     par RECEIVE_BLOCK_PROPOSAL, PROPAGATE, SYNC,
         COMMIT_PROPOSED_BLOCK, COMMIT_EMPTY_BLOCK in
        NODE [...] (1)
     ||
        NODE [...] (2)
     ||
        BOYCOTT -> MNODE [...] (3)
     ||
        BOYCOTT -> MNODE [...] (4)
     end par

  end process
	\end{verbatim}}

%
\subsection{CADP Noninterference Verification}
\label{sec:cadp}
%

With the LNT model above we can use the CADP toolbox to check equivalence-based properties. In particular,
for the purpose of noninterference analysis \linebreak we consider $\mathit{boycott}$ as the only high-level
action. When the number of malicious nodes is sufficiently large, we have that:
\cws{0}{\mathit{MAlgorand} \setminus \{ \mathit{boycott} \} \;\not\wbis{}\; \mathit{MAlgorand} \,/\, \{
\mathit{boycott} \}}
indicating that the presence of malicious nodes violates noninterference by making the outcome of the
protocol observably dependent on their behavior. Relation $\wbis{}$ is a $\tau$-abstracting equivalence such
as weak~\cite{Mil89a} or branching~\cite{GW96} bisimilarity.

To implement this equivalence check in CADP, we use a simple verification script written in SVL (Script
Verification Language)~\cite{GL01}. We start by generating from the \texttt{MALGORAND.lnt} file a labeled
transition system in the BCG (Binary Coded Graphs) format of CADP. We then produce
\texttt{restriction\_MALGORAND.bcg} and \texttt{hiding\_MALGORAND.bcg} -- the translations of
$\mathit{MAlgorand} \setminus \{ \mathit{boycott} \}$ and $\mathit{MAlgorand} \,/\, \{ \mathit{boycott} \}$
-- by using the SVL commands \texttt{hide} and \texttt{cut}. Lastly, we compare them with respect to weak
and branching bisimilarity. Here is the code of the script \texttt{MALGORAND.svl}, which effectively
implements the noninterference property called BSNNI~\cite{FG01,EABR25}:

	{\small\begin{verbatim}
  "MALGORAND.bcg" = generation of "MALGORAND.lnt"; 
  "restriction_MALGORAND.bcg" = cut "BOYCOTT" in "MALGORAND.bcg";
  "hiding_MALGORAND.bcg" = hide "BOYCOTT" in "MALGORAND.bcg";

  property WEAK_BSNNI is
     observational comparison "hiding_MALGORAND.bcg" == 
                              "restriction_MALGORAND.bcg";
     expected TRUE
  end property

  property BRANCHING_BSNNI is
     branching comparison "hiding_MALGORAND.bcg" == 
                          "restriction_MALGORAND.bcg";
     expected TRUE
  end property
	\end{verbatim}}

If in \texttt{MAIN} all nodes are honest, then the two aforementioned processes are equivalent:

	{\small\begin{verbatim}
  "restriction_MALGORAND.bcg" = cut "BOYCOTT" in "MALGORAND.bcg"
  "hiding_MALGORAND.bcg" = hide "BOYCOTT" in "MALGORAND.bcg"

  property WEAK_BSNNI
  PASS

  property BRANCHING_BSNNI
  PASS
	\end{verbatim}}

\noindent
whereas in the presence of sufficiently many malicious nodes the equivalence checks fail because the
malicious nodes can lead, after synchronizing on \texttt{BOYCOTT}, to commit the empty block:

	{\small\begin{verbatim}
  "restriction_MALGORAND.bcg" = cut "BOYCOTT" in "MALGORAND.bcg"
  "hiding_MALGORAND.bcg" = hide "BOYCOTT" in "MALGORAND.bcg"


  property WEAK_BSNNI
  FAIL

  property BRANCHING_BSNNI
  FAIL
	\end{verbatim}}

%
%
\section{Conclusions}
\label{sec:concl}
%
%

We have presented a process algebraic model of the Algorand BBA$^{*}$ protocol as described in~\cite{CM19}.
Our model captures the essential dynamics of committee-based consensus, including bit propagation, decision
conditions, and probabilistic committee selection through cryptographic self-sortition. The protocol steps
have been precisely formalized thanks to our fully synchronous abstraction.

We have then translated the pure model into the LNT language to enable verification with the CADP toolset,
in which we have implemented noninterference to assess the influence of malicious nodes attempting to
disrupt consensus. Our analysis has confirmed that a malicious minority cannot violate safety but can --
when exceeding the $1 / 3$ threshold -- cause the protocol to default to committing an empty block,
consistent with the design of BBA$^{*}$.

We plan to extend our investigation to a more realistic number of nodes -- by making the LNT specification
more modular and then resorting to compositional verification supported by CADP via SVL -- as well as to
different synchrony assumptions and richer threat models. We would like to apply our noninterference-based
methodology to the consensus protocol of other blockchains too.

This work contributes to the broader effort of applying formal methods to blockchain platforms. By embedding
into CADP the noninterference analysis techniques of~\cite{FG01,EABR25,EAB24,EAB25a,EAB25b} relying on
bisimilarity checking, we have paved the way to automated security verification for a variety of different
systems that can include quantitative aspects or revert their computations.

\medskip
\noindent
\textbf{Acknowledgment.} This research has been supported by the PRIN 2020 project \emph{NiRvAna --
Noninterference and Reversibility Analysis in Private Blockchains}.

\bibliographystyle{splncs04}
\bibliography{biblio}

\end{document}